# Evaluation of Petro Physical Properties of the Reservoirs in XY Field Gulf of Mexico (GOM)

Oko, Christian Obinna[1], Igbo, Nkechinyere Elem[2], Igbo, Michael Elem[3*], and Kalu, Jonah[4].
[1,2,3,4]Department of Science Laboratory Technology, Federal Polytechnic Afikpo, Ebonyi State, Nigeria.

**Abstract:-** A composite log suite that comprised of gamma ray, resistivity, density and neutron logs of six wells (Agate, Diamond, Apatite, Calcite, Copper and Jasper) were employed to evaluate the petrophysical properties of the reservoirs of interest in XY field Golf of Mexico. This was done to evaluate the trend and values of the petrophysical properties of the reservoirs in the field, while the objective was to delineate and predict the reservoir system quality and performance.

Correlation was carried out across the 6 wells using gamma ray, resistivity, neutron and density logs. The distribution and occurrence of these lithostratigraphic units appeared to reflect the influence of basin morphology and sea level variation. Tops and bases of 6 reservoirs of interest were mapped. Shale Volumes (Vsh) were computed using gamma ray log; the reservoir versus non-reservoir was delineated by applying cut-off of 80% computed volume of shale with the guide of spontaneous potential (SP) and deep resistivity logs for wells Agate and Diamond.

Computed Vsh was used for effective porosity to discount for the effects of clay bound water. The average porosity values ranged from 15 – 35% and 10 – 30%, Permeability averages of about 43mD – 75mD and 45mD – 118mD, and water saturation values averaging from 14.7% - 76.6% and 13.8% - 82% were gotten across the reservoirs in wells Agate and Diamond respectively.

## I. INTRODUCTION

The study area, XY field Gulf of Mexico, is an ocean basin that lies between North American plate and the Yucatan block. Its depocenter consists of sediments that have been deposited in time from the Jurassic period through to the Holocene period with thickness of about 20km [3]. Sediments from the North American continent filled almost half of the basin, primarily by offlap of the northern margin and northwestern margin. The basin currently has an abyssal plain that lies at a depth of 3km [4].

In the east, the Gulf floor is dominated by morphology akin to the Late Quaternary Mississippi fan while the northern Gulf margin shows a bathymetrically complex morphology that terminated abruptly in the Sigsbee escarpment in the west and merged into the Mississippi fan in the east [14]. In the west, the Gulf margin shows intermediate width that is bathymetrically complex. In recent times, the shelf edge reflects a well-defined increase in basinward gradient lying generally at a depth of about 100–120 m. The northern, eastern and northwestern Gulf of Mexico is bounded by broad and low-gradient shelves with thickness that range from 100 to 300 km in the landward direction. Presently, the Florida and Yucatan platforms, which bound the Gulf basin on the east and south, persist as sites of carbonate deposition.

Onshore, the north and northwest Gulf margin shows broad coastal plain. At the lower coastal plain, Neogene and Quaternary strata underlies the flat and low-relief surface plain while the upper plain shows a modest relief that is less than 100m created by the incision of Quaternary strata into older Neogene, Paleogene, and Late Cretaceous strata through numerous small and large rivers. Various sediments from Cenozoic, Mesozoic, and remnant Paleozoic uplands which includes, the Lower Cretaceous limestone-capped Edwards Plateau, the Sierra Madre Oriental of Mexico, Ouachita Mountains of southern Arkansas, the Trans-Pecos mountains of West Texas, and the Cumberland Plateau and southern Appalachian Mountains of northern Mississippi and Alabama, bounds the Gulf basin [5].

The northeastern Gulf basin merged into the south Atlantic coastal plain across northern Florida with the structural basin boundary generally placed very close to the current west coast of the Florida peninsula [8].

## II. MATERIALS

Materials used for this research work includes PETREL software, TECHLOG software, ECLIPSE software and other research and training materials.

### 2.1. DATA INVENTORY
Summary of dataset used for the purpose of this research work is presented on table shown below;

| Input Data Set Requirement for Geological/Geophysica/Petrophysical Evaluation Studies | | | | | | | | | | | | | | | |
|---|---|---|---|---|---|---|---|---|---|---|---|---|---|---|---|
| Data Set Type | | | | | | Wellogs | | | | | | | | | |
| NAMEWELL | CHECKSHOT | CALI | CILD | DPHI | DRHO | DT | GR | ILD | ILM | NPHI | PHIT | RHOB | RT | SFLU | SP | VCL |
| AGATE | YES | YES | YES | YES | YES | YES | YES | YES | YES | YES | YES | YES | YES | YES | YES | YES |
| APATITE | YES | NO | YES | YES | NO | YES | YES | YES | NO | YES | YES | NO | YES | YES | YES | YES |
| CALCITE | YES | NO | YES | NO | NO | YES | YES | YES | NO | YES | NO | YES | YES | YES | YES | YES |
| COPPER | YES | YES | YES | NO | YES | YES | YES | YES | NO | NO | YES | YES | YES | NO | YES | YES |
| DIAMOND | YES | YES | YES | YES | YES | YES | YES | YES | YES | YES | YES | YES | YES | YES | YES | YES |
| JASPER | YES | NO | YES | NO | NO | YES | YES | YES | NO | NO | YES | NO | YES | NO | YES | YES |
| | | | | | | | | | | | | | | | | |
| | Seismic | YES | | | | | | | | | | | | | | |
| | PVT | NO | | | | | | | | | | | | | | |
| | Tops | YES | | | | | | | | | | | | | | |
| | Well test | NO | | | | | | | | | | | | | | |
| | Core Logs | NO | | | | | | | | | | | | | | |
| | Capillary Pressure Data | NO | | | | | | | | | | | | | | |

Table 1. *Data Inventory Table for the six wells of interest*

➢ **WELL LOG DATA**

All the wells had caliper (CILD), Sonic (DT), Gamma Ray (GR), resistivity (ILD), porosity (PHIT), spontaneous potential (SP), and clay volume (VCL). Four wells had fluid saturation (SFLU) and three wells had caliper (CAL), density porosity (DPHI), density (RHOB), density correction (DRHO), resistivity (ILM), neutron (NPHI).

➢ **GEOLOGICAL SETTING**

Gulf of Mexico Basin is ranked among the world's great petroleum mega-provinces, with hydrocarbon production history of more than 100 years. It is a rift basin located between North America and the Yucatan Block. The field is adjacent to the Caribbean Sea and surrounded by Eastern Mexico, Texas, the Southeastern Gulf States of the United States, and Cuba.

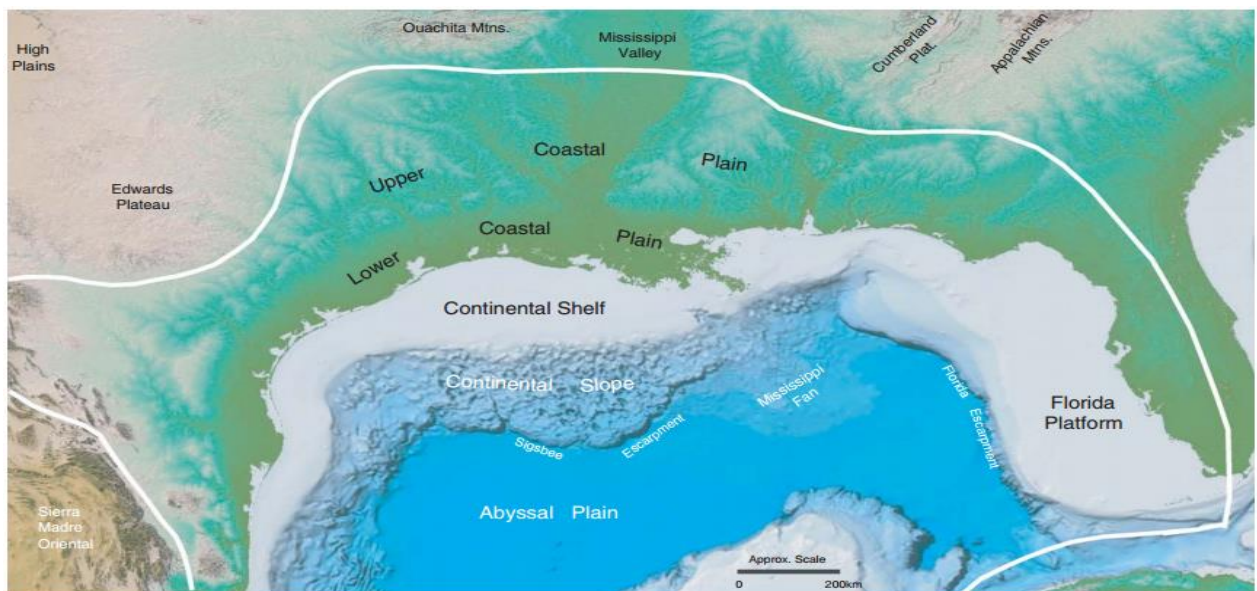

**Figure 1.** *Physiographic elements of the Gulf basin, showing adjacent North America. The white outline shows approximate geological limits of the basin.*

➢ **WELL LOCATION**

The six wells of interest are located on the Texas-Louisiana shelf. The shelf has a thickness of about 100km off the Rio Grande in Texas and more than 200km in the southern part of Texas-Louisiana boundary. Large amounts of salt sedimentary sequence in the northwestern part of the Gulf shelf is the prominent structural feature in the area. The imminent effect of the salt in the area is the high tectonic movement of the shelf caused principally by the salt depositional sequence.

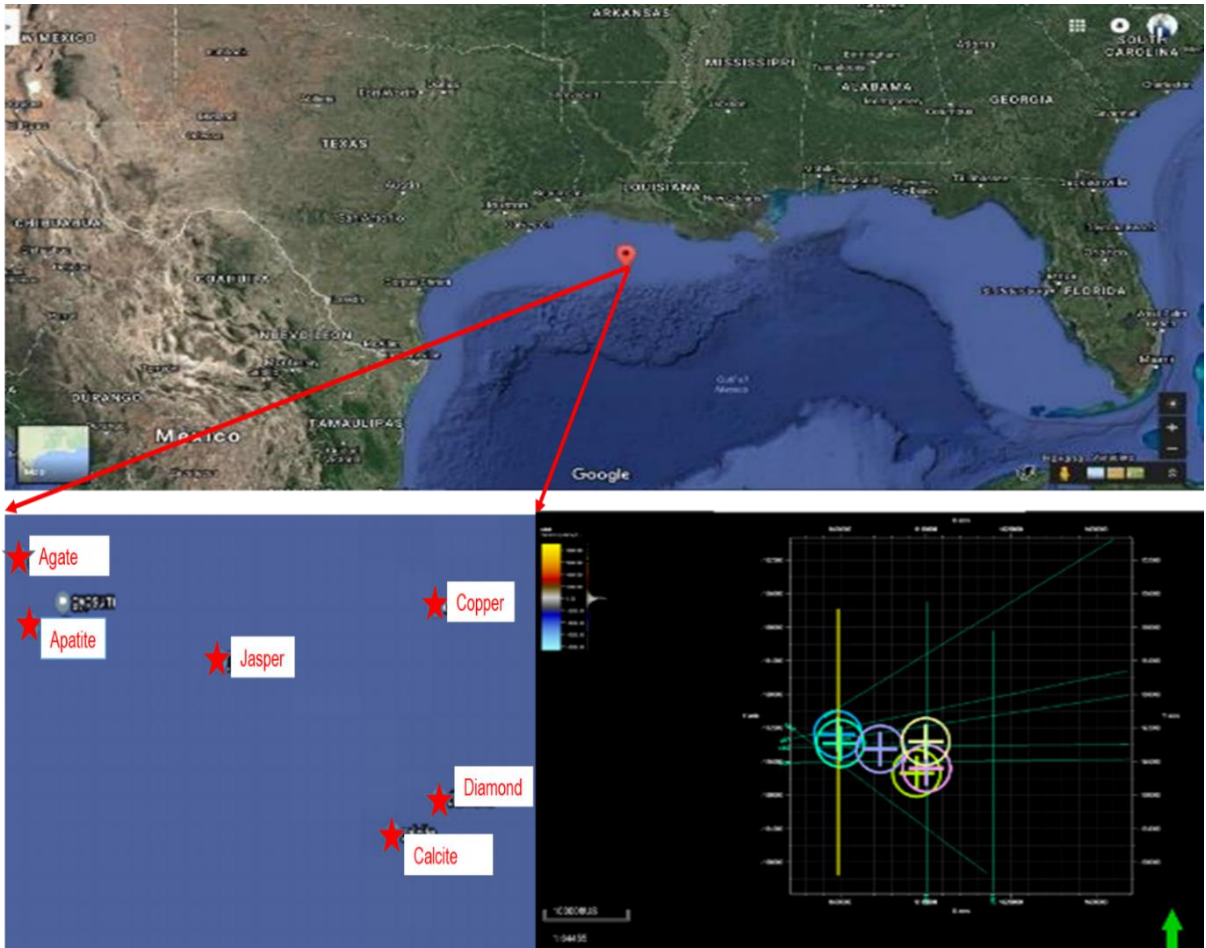

**Figure 2.** *(a) Showing the wells location in the GoM (b) Showing enlarged view of the wells in the study location. (C) Showing wells location on a 2D Petrel window*

## III. METHOD

An integrated approach was applied in this study. The knowledge from the various subsurface disciplines was employed to develop 3D Reservoir Model of the field with the help of Schlumberger Oilfield Services Software (PETREL, TECHLOG AND ECLIPSE) in Portharcourt. Interdisciplinary reviews and interpretations were provided at each milestone of this project to ensure consistency in the geological, geophysical and reservoir engineering concepts of the study.

The processes employed in the study include:
- Data load, QA/QC, and data conditioning
- Sonic log calibration (TDR generation)
- Synthetic generation
- Well-to-seismic tie
- Seismic interpretation (fault and horizon interpretation)
- Generating surfaces for the reservoirs of interest
- Velocity modeling
- Time to Depth conversion
- Structural modeling of faults and horizons
- 3D grid model building (static), zoning and layering of the grids
- Property modeling
- Data analysis

### 3.1. STRUCTURAL MODELING

A structural model is regarded as the skeleton or container of the subsurface. It is the geological model of the area otherwise called the Geomodel [7]. To model any geologic subsurface, the modeler starts with a sound structural framework of the area which represents the relation between the faults and the seismic horizons showing the important layers of interest [12]. Structural modeling in Petrel involves Pillar Gridding, Fault Modeling, and Horizon picking (Layering).

**Fault Modeling:** The first step in fault modeling with petrel involves defining the faults of interest [1]. This helps to define the shape of each fault that must be modeled. The process was achieved by generating "Key Pillars".

**Pillar Gridding:** Pillar gridding involves making grids based on the already defined faults. In this stage, a set of pillars was inserted in between interpreted faults and in every corner of the grids in the entire project area. The result of this process is a "Skeleton grid", defined by all the faults and all the pillars.

**Layering:** After the fault modeling and the pillar gridding process, horizons were inserted into the faulted 3D grid. The 3D grid was converted to time/depth by associating it with time/depth maps and well tops. After inserting the horizons, the zones of interest (based on geological input such as isochors) were inserted, and the final step was to make fine-scale layerings suitable for property modeling.

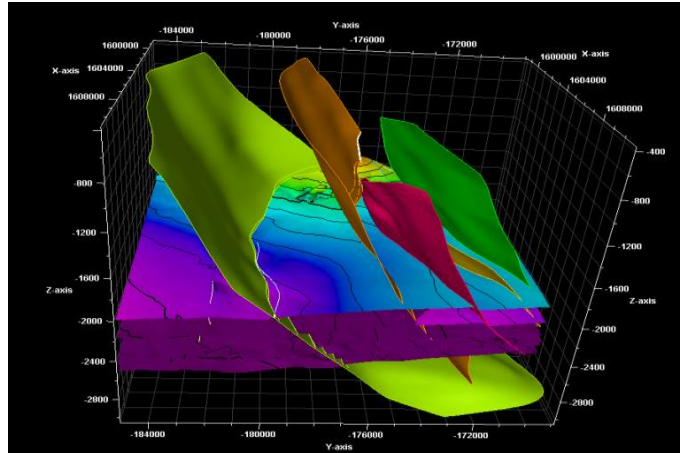

*Figure 3. Fault and Horizon structural frameworks*

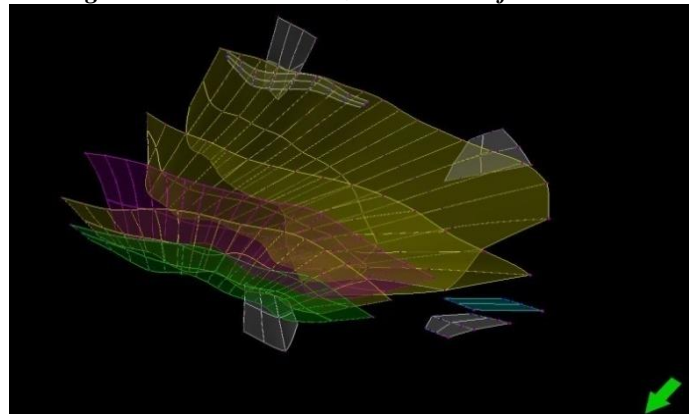

**Figure 4.** *Pillar grid fault framework*

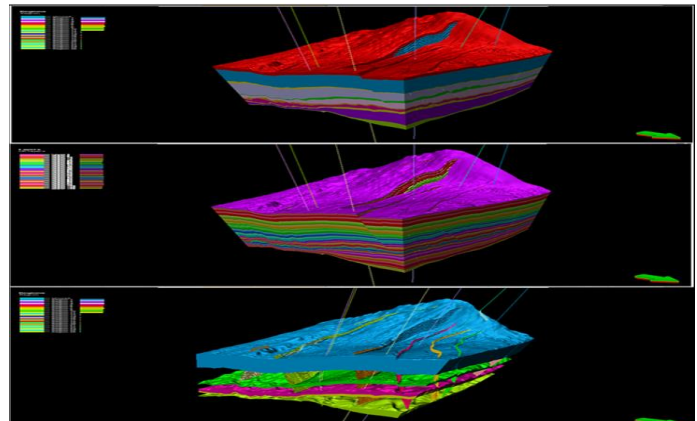

**Figure 5.** *Layered structural model*

> *Property Modeling*

Property modeling involves filling the 3D grid cells with discrete/continuous properties. The goal is to achieve a realistic property model by using all geological information available to build the model [13]. The Property modeling of the field was done with Petrel software and the process was split into three steps:

**Geometrical modeling:** This involves using some pre-defined functions to generate properties such as Bulk Volume, Depth, Height above Contact, and more [2]. These properties are very important in volumetric evaluations and in mathematical operations between petrophysical properties (e.g. for Sw transforms).

**Facies modeling:** This involved the population of discrete data such as lithofacies, into 3D grid cells. This helps to give true understanding of the geological processes, and to capture facies architectures like reservoir connectivity, high level of heterogeneity and to confirm descriptive facies information such as shape, size, orientation, proportion, distribution and statistics.

**Petrophysical modeling:** This involves the interpolation and simulation of continuous data such as porosity, permeability and saturation into the 3D grid cells [9]. The petrophysical properties vary from facies to facies i.e. different petrophysical property distributions in different facies with spatial variations for each petrophysical parameter.

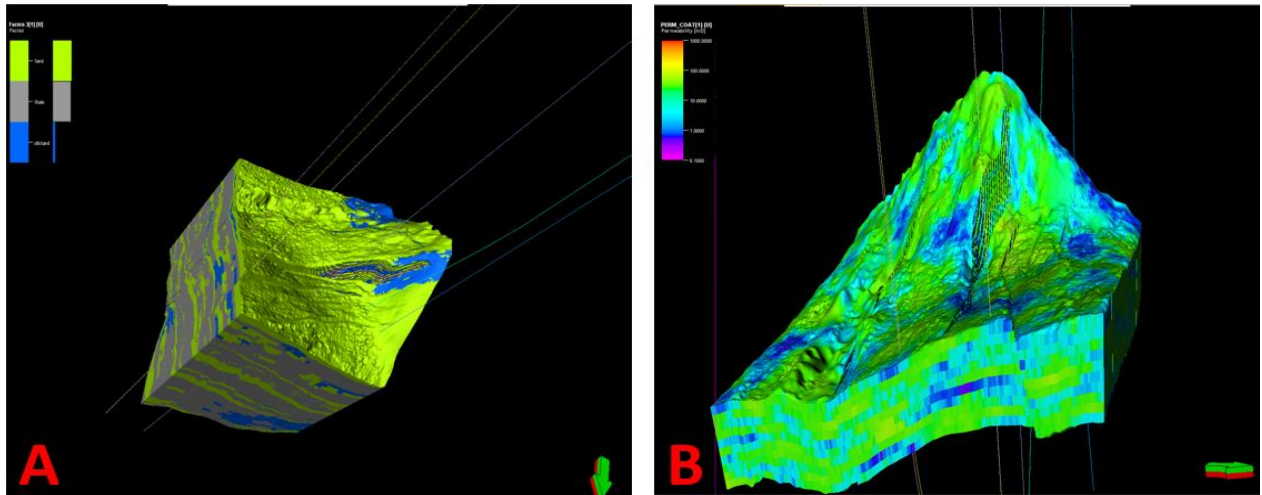

**Figure 6.** *3D Grid showing (A) facies distribution and (B) Porosity distribution*

➤ *Model Upscaling*

Model upscaling involves merging the fine cells of the 3D Grid to obtain a coarse model while still preserving the geologic features important to the reservoir flow dynamic. For this study a 4:1 flow-based upscale was performed to retain important flow characteristics of the fine-scale model. This was done to generate a stable and fast simulation model which is important and critical to allow the engineers spend enough time interpreting results and testing more scenarios.

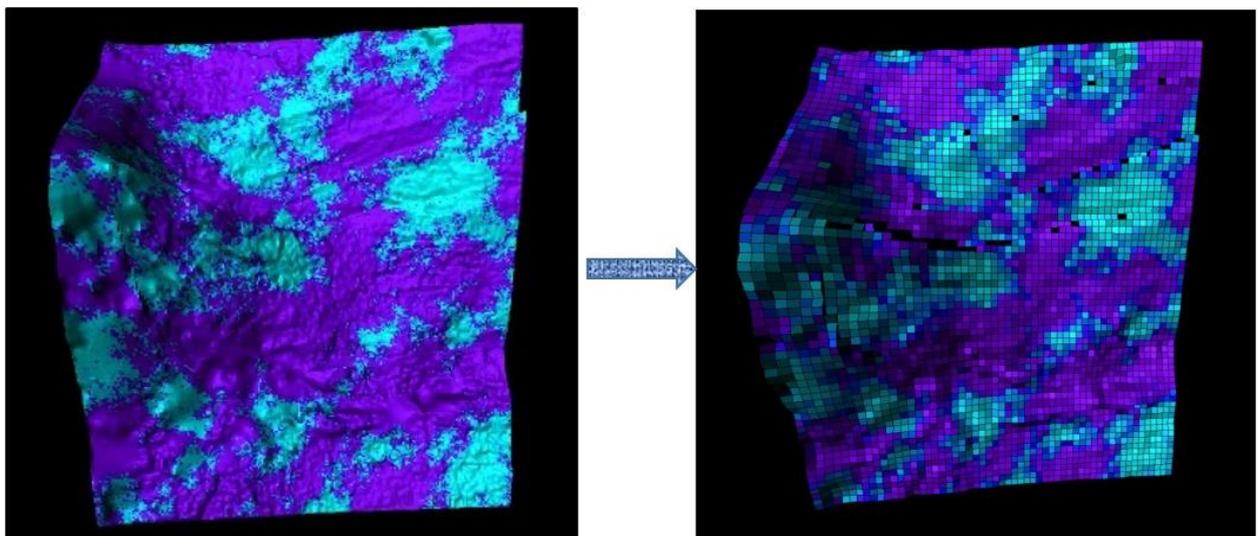

**Figure 7.** *Showing fine scale and coarse scale 3D grid geomodel*

## IV. RESULTS AND DISSCUSSIONS

### 4.1. Petrophysical Evaluation

Petrophysical evaluation carried out on the measured log using the resistivity (ILD), gamma ray (GR), and crossplot of neutron (NPHI) and density (RHOB) logs indicated the presence of hydrocarbon within intervals of high resistivity and neutron density crossovers. This is the case in the two control wells, Agate and Diamond wells as can be seen in figure 8, below. Due to lack of neutron and density logs on wells Apatite, Calcite, Copper and Jasper, conclusions could not be made on the presence of hydrocarbons on the different reservoir intervals.

Average porosity across the reservoirs is 0.30 - 0.40. Kobe reservoir contains gas while the other reservoir contains oil.

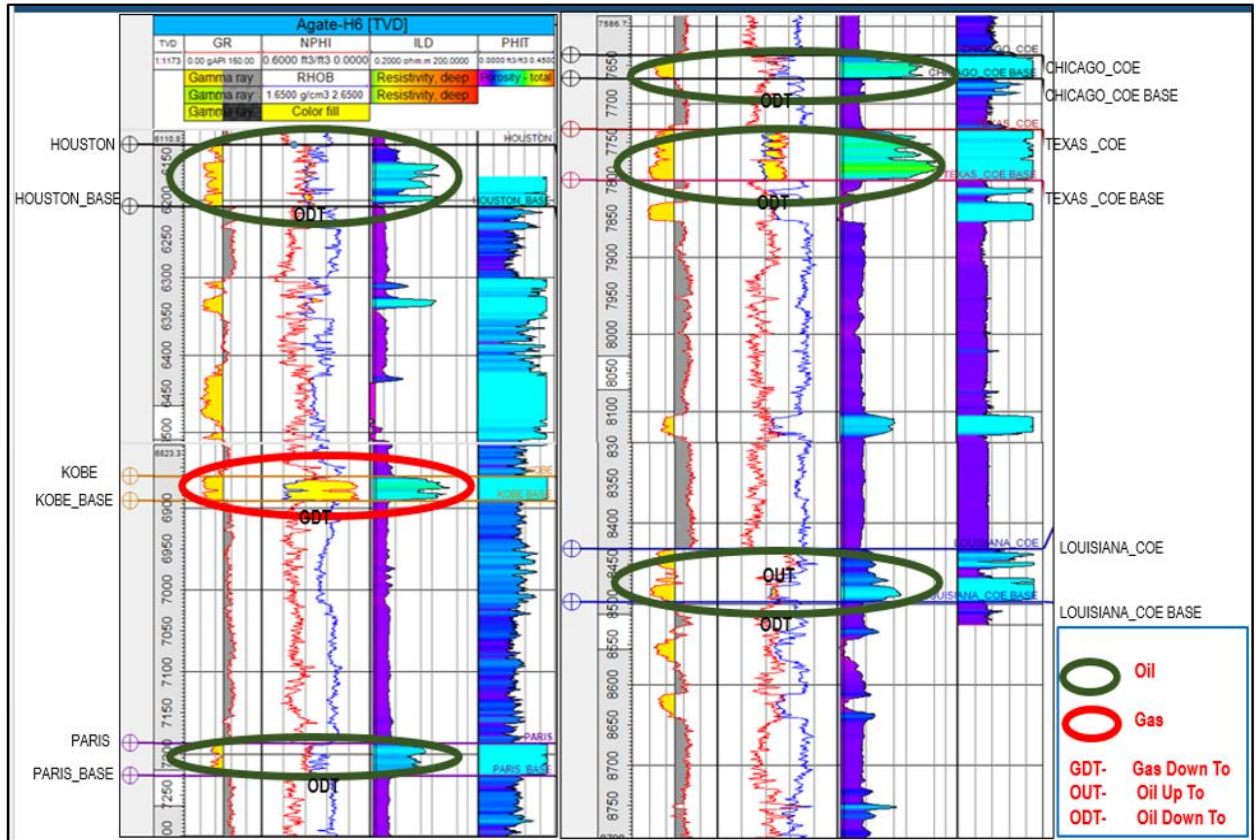

**Figure 8.** *Result showing reservoirs of interest and hydrocarbon type throughout the column of well Agate*

### 4.2. Petrophysical Properties

Shale Volumes (Vsh) were computed using gamma ray log; the reservoir versus non-reservoir was delineated by applying cut-off of 80% computed volume of shale with the guide of spontaneous potential (SP) and deep resistivity logs for wells Agate and Diamond. Computed Vsh was used to effective porosity to discount for the effects of clay bound water. The average porosity values ranged from 15 – 35% and 10 – 30%, Permeability averages of about 43mD – 75mD and 45mD – 118mD, and water saturation values averaging from 14.7% - 76.6% and 13.8% - 82% were gotten across the reservoirs in wells Agate and Diamond respectively.

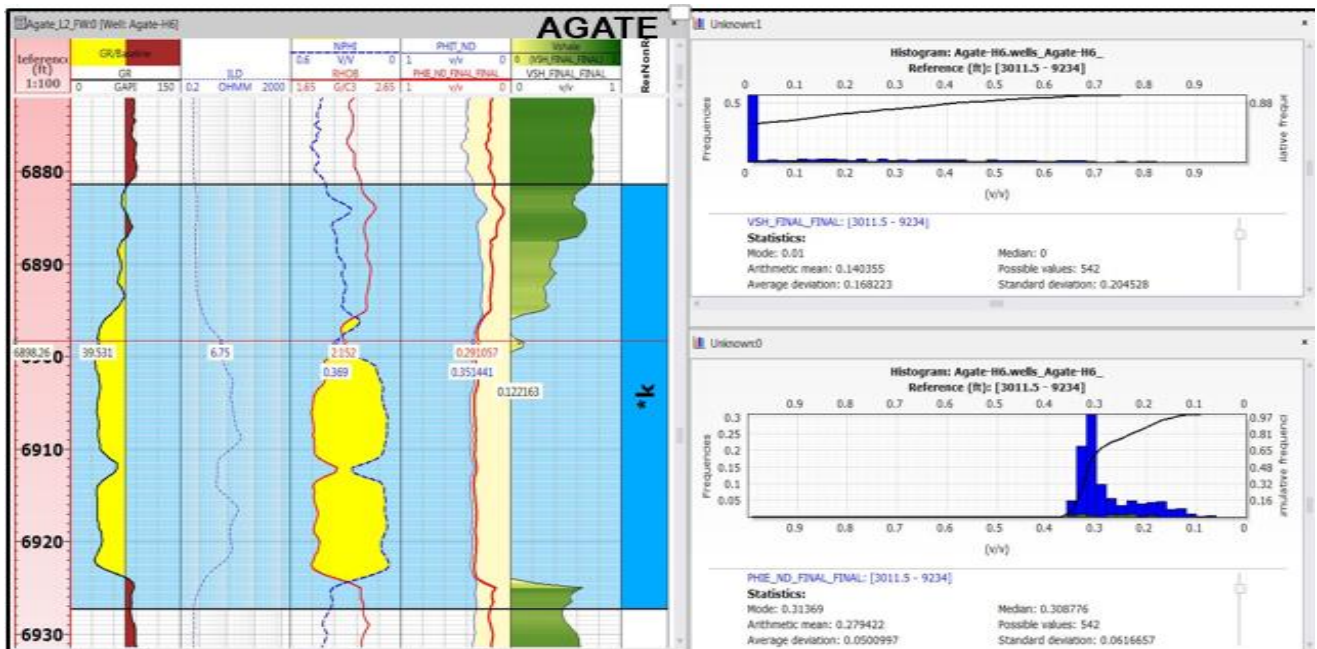

**Figure 9.** *Result of the shale volume and effective porosity curves generated for well Agate*

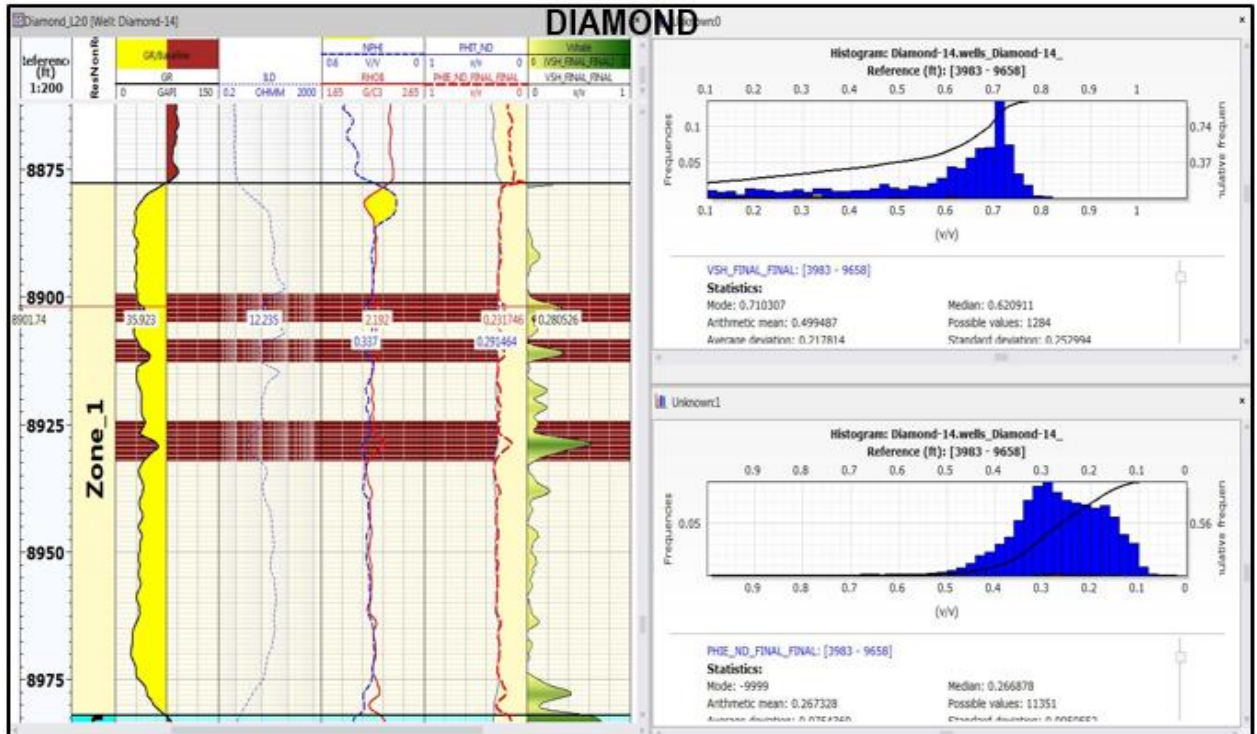

**Figure 10.** *Result of the shale volume and effective porosity curves generated for well Diamond*

| | Well | Top | Bottom | Reference unit | Gross | Net | Net to Gross | Av_Shale Volume | Av_Porosity | Av_Water Saturation |
|---|---|---|---|---|---|---|---|---|---|---|
| 1 | Agate-H6 | 6089.025 | 6183.049 | ft | 94.023 | 2.5 | 0.027 | 0.261 | 0.374 | 0.806 |
| 2 | Agate-H6 | 6183.049 | 6241.637 | ft | 58.588 | 49 | 0.836 | 0.276 | 0.366 | 0.427 |
| 3 | Agate-H6 | 6338.333 | 6378.569 | ft | 40.235 | 16.5 | 0.41 | 0.245 | 0.358 | 0.484 |
| 4 | Agate-H6 | 6458.813 | 6542.802 | ft | 83.989 | 3.5 | 0.042 | 0.27 | 0.347 | 0.819 |
| 5 | Agate-H6 | 6542.802 | 6893.528 | ft | 350.726 | 13 | 0.037 | 0.272 | 0.356 | 0.655 |
| 6 | Agate-H6 | 6893.528 | 6925.467 | ft | 31.938 | 30.472 | 0.954 | 0.231 | 0.343 | 0.384 |
| 7 | Agate-H6 | 6893.528 | 6925.467 | ft | 31.938 | 27.5 | 0.861 | 0.199 | 0.351 | 0.333 |
| 8 | Agate-H6 | 7220.541 | 7265.4 | ft | 44.859 | 37.5 | 0.836 | 0.317 | 0.343 | 0.528 |
| 9 | Agate-H6 | 7684.871 | 7721.7 | ft | 36.829 | 31.2 | 0.847 | 0.345 | 0.321 | 0.392 |
| 10 | Agate-H6 | 7721.7 | 7785.095 | ft | 63.395 | 0.3 | 0.005 | 0.714 | 0.275 | 0.791 |
| 11 | Agate-H6 | 7785.095 | 7918.106 | ft | 133.012 | 68 | 0.511 | 0.276 | 0.349 | 0.267 |
| 12 | Agate-H6 | 8184.33 | 8214.244 | ft | 29.914 | 27.5 | 0.919 | 0.321 | 0.337 | 0.48 |
| 13 | Agate-H6 | 8528.502 | 8754.707 | ft | 226.205 | 34.5 | 0.153 | 0.211 | 0.324 | 0.556 |
| 14 | Diamond-14 | 8876.618 | 8983.662 | ft | 107.044 | 104.382 | 0.975 | 0.267 | 0.293 | 0.357 |
| 15 | Diamond-14 | 8876.618 | 8983.662 | ft | 107.044 | 101 | 0.944 | 0.254 | 0.294 | 0.338 |
| 16 | Diamond-14 | 9233.857 | 9300.98 | ft | 67.123 | 65.623 | 0.978 | 0.488 | 0.309 | 0.88 |
| 17 | Diamond-14 | 9233.857 | 9300.98 | ft | 67.123 | 22 | 0.328 | 0.356 | 0.315 | 0.658 |
| 18 | Diamond-14 | 9456.169 | 9575.418 | ft | 119.249 | 83.331 | 0.699 | 0.538 | 0.282 | 0.766 |
| 19 | Diamond-14 | 9456.169 | 9575.418 | ft | 119.249 | 38.5 | 0.323 | 0.403 | 0.294 | 0.53 |

**Table 2.** *Petrophysical evaluation table (reservoir sums and averages for wells Agate and Diamond)*

## V. CONCLUSION

In conclusion, this research work evaluated the petrophysical properties of the reservoir sand bodies of XY field Gulf of mexico (GOM). Six wells of interest were picked for this study. Existing seismic, together with available well logs were interpreted. Thus, the Horizons (layer tops) representing the six reservoirs of interest were picked while faults were interpreted to define the storage container boundaries. Structural and petrophysical analysis of the field was done and the results from the study shows that the average porosity values of the field ranged from 15% - 35% and 10% - 30% in wells Agate and Diamond Permeability values of 43mD – 75mD and 45mD – 118mD, and water saturation values averaging from 14.7% - 76.6% and 13.8% - 82% were gotten across the reservoirs in wells Agate and Diamond respectively.

Due to lack of neutron and density logs on wells Apatite, Calcite, Copper and Jasper, conclusion could not be made on the petrophysical properties of the wells on the different reservoirs intervals.